\documentclass{emulateapj}
\usepackage{graphics,graphicx}
\usepackage{psfig}
\usepackage{times}
\usepackage{comment}

\providecommand{\mb}{$\Delta m_{15}(B)$}

\providecommand{\mu}{$\Delta m_{15}(U)$}

\begin{document}
\title{Reddened, Redshifted, or Intrinsically Red?\\
 Understanding Near-Ultraviolet Colors \\
of Type I\lowercase{a} Supernovae}


\author{Peter~J.~Brown\altaffilmark{1}, Nancy J. Landez\altaffilmark{1}, Peter A. Milne\altaffilmark{2}, \& Maximilian D. Stritzinger\altaffilmark{3} }

\altaffiltext{1}{George P. and Cynthia Woods Mitchell Institute for Fundamental Physics \& Astronomy, 
Texas A. \& M. University, Department of Physics and Astronomy, 
4242 TAMU, College Station, TX 77843, USA; 
pbrown@physics.tamu.edu}            
\altaffiltext{2}{Steward Observatory, University of Arizona, 933 N. Cherry Avenue, Tucson, AZ 85721, USA 0000-0002-0370-157X}

\altaffiltext{3}{Department of Physics and Astronomy, Aarhus University, Ny Munkegade 120, DK-8000 Aarhus C, Denmark}

\begin{abstract}

Understanding the intrinsic colors of Type Ia supernovae (SNe Ia) is important 
to their use as cosmological standard candles.  Understanding the effects of reddening and redshift on the observed colors are complicated and dependent on the intrinsic spectrum, the filter curves, and the wavelength dependence of reddening.  We present ultraviolet and optical data of a growing sample of SNe Ia observed with the Ultra-Violet/Optical Telescope on the Swift spacecraft and use this sample to re-examine the near-UV (NUV) colors of SNe Ia. 
We find that a small amount of reddening ($E(B-V)=0.2$ mag) could account for the difference between groups designated as NUV-blue and NUV-red, and a moderate amount of reddening ($E(B-V)=0.5$ mag) could account for the whole NUV-optical differences.   The reddening scenario, however, is inconsistent with the mid-UV colors and color evolution.  The effect of redshift alone only accounts for part of the variation.  Using a spectral template of SN2011fe we can forward model the effects of redshift and reddening and directly compare with the observed colors.  We find that some SNe are consistent with reddened versions of SN2011fe, but most SNe Ia are much redder in the $uvw1-v$ color than SN2011fe reddened to the same $b-v$ color.  
The absolute magnitudes show that two of five NUV-blue SNe Ia are blue because their near-UV luminosity is high, and the other three are optically fainter.   We also show that SN~2011fe is not a ``normal'' SN Ia in the UV, but has colors placing it at the blue extreme of our sample.

\end{abstract}


\section{Introduction \label{intro}}

Type Ia Supernovae (SNe Ia) are important cosmological tools because their high optical luminosity correlates tightly with the optical light curve shape and color \citep{Pskovskii_1977,Phillips_1993,Riess_etal_1996_MLCS,Phillips_etal_1999}, allowing them to be used as standard candles \citep{Branch_1998,Leibundgut_2001}.  The near-infrared luminosity is found to have a smaller dependence on light curve shapes while also being less susceptible to the effects of interstellar extinction \citep{Krisciunas_etal_2000,Wood-Vasey_etal_2008,Mandel_etal_2011,Kattner_etal_2012}.  In contrast, the ground-based near-ultraviolet has been found to be less homogeneous (e.g. \citealp{Jha_etal_2006_U,Kessler_etal_2009}), likely a combination of intrinsic differences such as metallicity and ejecta velocity \citep{Burns_etal_2014}, interstellar extinction, and the effects of different filter shapes and atmospheric absorption \citep{Krisciunas_etal_2013}.  

The launch of the Swift spacecraft \citep{Gehrels_etal_2004} with its Ultra-Violet Optical Telescope (UVOT; \citealp{Roming_etal_2005}) ushered in a new era of SN observations in the UV (see \citealp{Brown_etal_2015_10} for a review).  Data from Swift/UVOT revealed an increased dispersion in the near and mid-UV wavelengths \citep{Brown_etal_2010,Milne_etal_2010}.  \citet{Milne_etal_2013} suggested the possibility that this dispersion could be the result of two distinct classes separated by a shift in their NUV-optical color evolution, with ``NUV-red'' SNe Ia being the majority and ``NUV-blue'' SNe Ia comprising about 25\% of the SNe Ia. This came from a sample of 23 SNe with Swift/UVOT photometry and 33 SNe with HST near-UV spectrophotometry.  \citet{McClelland_2013} noted a similar difference in the fraction of the luminosity coming from the UV in a subset of the same Swift SNe Ia.  The cause of the difference is unknown, with velocity, density gradient \citep{Sauer_etal_2008}, mixing in the ejecta \citep{Hoeflich_etal_1998}, and metallicity  \citep{Sauer_etal_2008,Walker_etal_2012} differences being possibilities.  Additionally, the detection of unburned carbon is much more common amongst NUV-blue SNe Ia  \citep{Thomas_etal_2011,Milne_etal_2013}, though non-uniformity in the spectroscopic follow-up and analysis of SNe Ia make quantification difficult.  Whether they represent distinct populations or are a continuum is also uncertain.  \citet{Milne_etal_2015} found that similar groupings can be identified in the spectrophotometry of near-UV spectra of 59 higher redshift SNe Ia.
Disturbing for those relying on the homogeneous nature of SNe Ia is the finding that the relative fractions of NUV-blue and NUV-red SNe Ia changes with redshift.  

One difficulty with a classification based on color is the uncertain effect of reddening from circumstellar or interstellar dust.  \citet{Milne_etal_2013} compared the peak colors to reddening vectors and found that the whole sample of colors could not result from a single intrinsic color modified by either a Milky Way (MW) dust law (R$_V$=3.1, \citealp{Cardelli_etal_1989}) or a circumstellar scattering dust law (R$_V$=1.7, \citealp{Wang_2005,Goobar_2008}). Separate samples were compared spectroscopically in \citet{Milne_etal_2013} and \citet{Milne_etal_2015} with several pairs showing a distinct block of emission in the NUV-blue SNe Ia as opposed to the continuum expected from extinction differences.  \citet{Cinabro_etal_2016}, however, find that a model with a distribution of colors, brightness, stretch and extinctions better matches the higher redshift photometric samples from SDSS-II and SNLS than a model with a bimodal distribution in the near-UV colors.  Their assumed spectroscopic model creating the near-UV differences, however, is not consistent with the spectroscopic or photometric differences seen \citep{Milne_etal_2013,Milne_etal_2015}.  Clearly a better understanding of the near-UV differences is needed.  

Here we study in more detail the effects of color dispersion and reddening within an expanded sample of nearby SNe Ia. In Section \ref{observations} we detail the sample we presently use.  Section \ref{results} compares the sample to the \citet{Milne_etal_2013} color differences and examine the color and color-color  evolution and absolute magnitudes.  We discuss the consequences of our findings in Section \ref{discussion} and summarize in Section \ref{summary}.

 
\begin{figure*} 
\plottwo{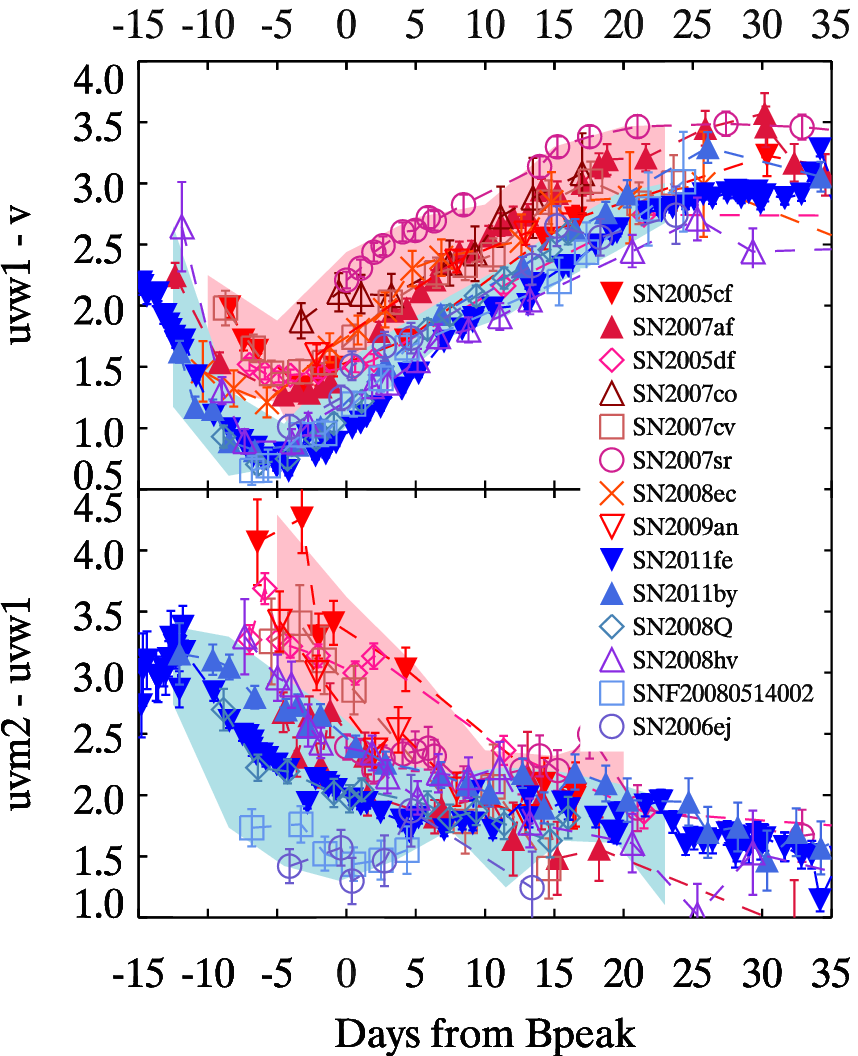}{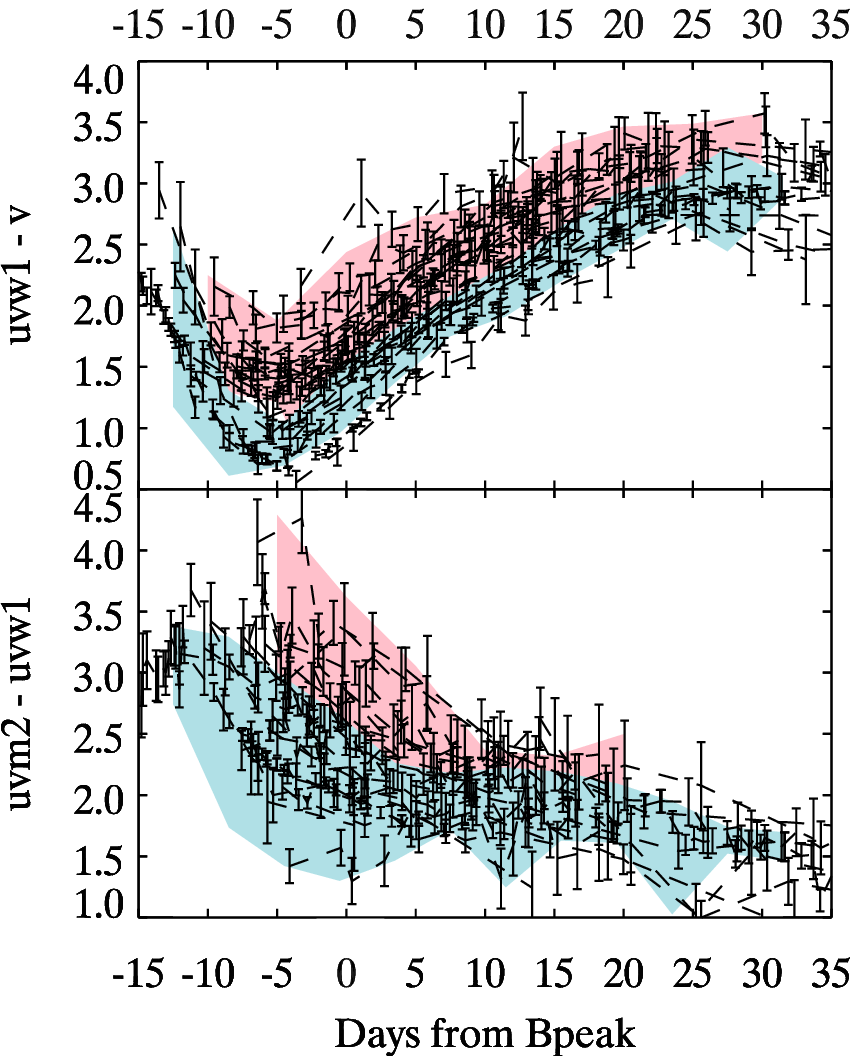}  
\caption[Results]
        {Left:  Fourteen SNe Ia meeting our criteria from the sample of \citet{Milne_etal_2013} are used to define the NUV-blue and NUV-red color evolution curves.   
Right: The expanded UVOT SNe Ia sample (31 SNe) considered in this work.
 } \label{fig_milne}    
\end{figure*} 


\begin{figure*} 
\plottwo{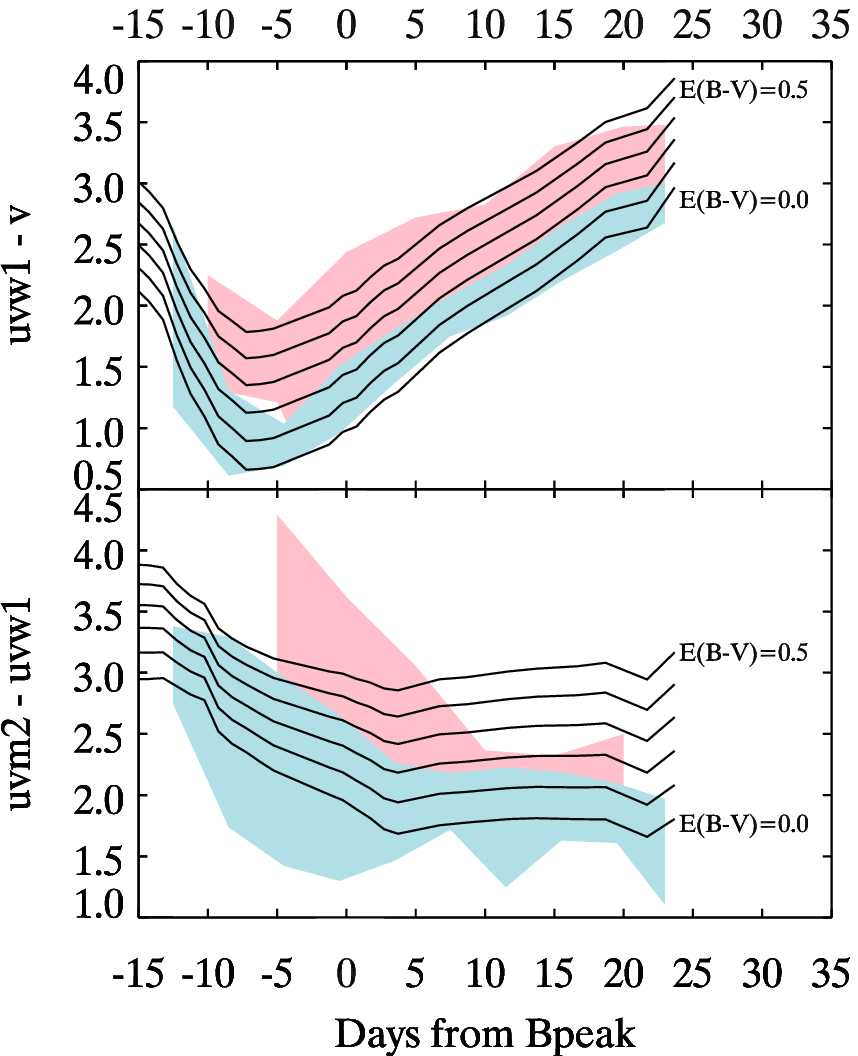}{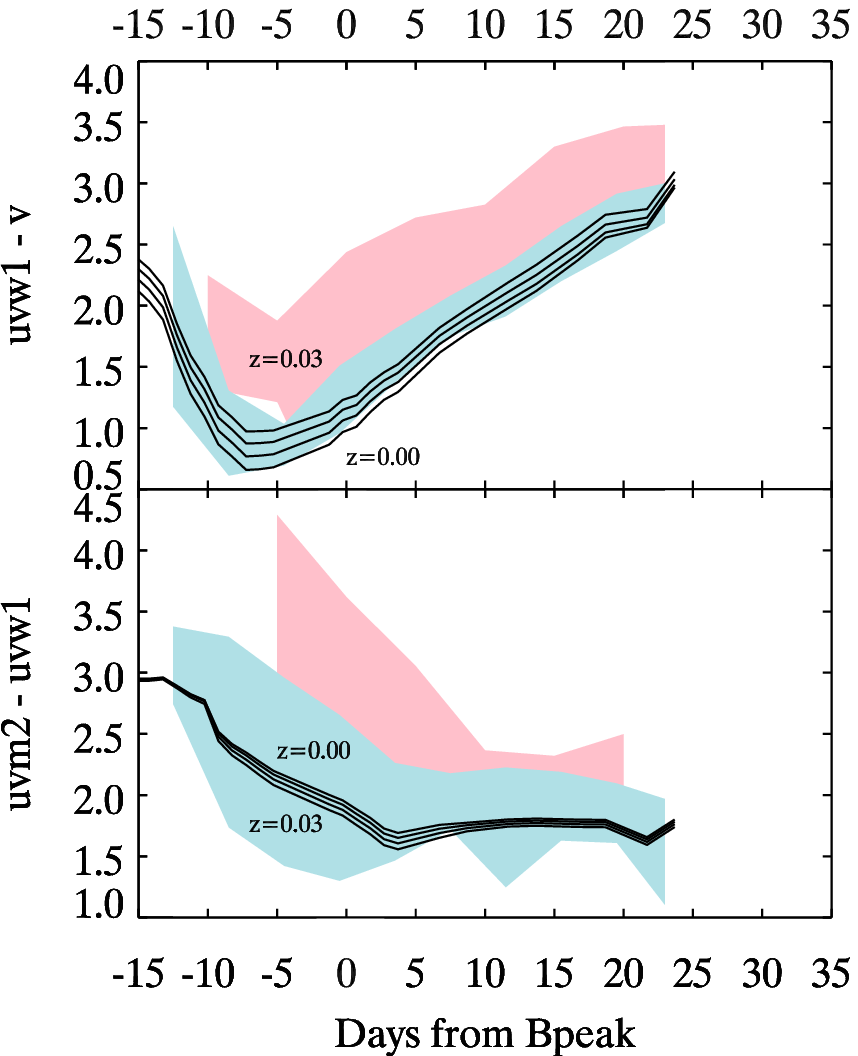}  
\caption[Results]
        {
Left: The color evolution curves of the SN~2011fe template are shown after the application of a MW (R$_V$=3.1) reddening law in steps of E(B-V)=0.1 mag.  The colored regions correspond to the range of colors seen for SNe Ia identified as NUV-blue and NUV-red in \citet{Milne_etal_2013} and shown in Figure \ref{fig_milne}.  
Right: The color evolution curves of the SN~2011fe template are shown after redshifting in steps of z=0.01.

 }\label{fig_redz}    
\end{figure*}

\section{Observational Sample} \label{observations}
 
 Swift/UVOT has observed over 500 SNe of all types in its twelve years of operation (see \citealp{Brown_etal_2015_10} for a review of the first ten years).  Most of the observations use the six UV and optical filters.  The $uvw2$ filter has a  central wavelength of 1941 \AA\ and an effective wavelength (using the type Ia SN~1992A spectrum; \citealp{Kirshner_etal_1993}) of 3064 \AA.  The equivalent central and effective wavelengths for the other filters are $uvm2$: 2248 and 2360, $uvw1$: 2605 and 3050, $u$: 3464 and 3600, $b$: 4371 and 4340, and $v$: 5442 and 5400 \AA.  These numbers do not capture the full complexity of the filter throughputs and the effective transmission for a SN Ia spectrum; the reader is referred to \citet{Brown_etal_2010} and \citet{Brown_etal_2016} for further details.

The Swift SN sample has grown considerably since \citet{Milne_etal_2013}, warranting a re-examination of the sample's bulk properties.   \citet{Milne_etal_2013} used 23 SNe Ia, of which 14 would meet the \mb, subclassification, and light curve quality cuts we use below to arrive at our sample of 29 SNe Ia.  The correlation between NUV flux and the optical decline rate parameter \mb~has been seen with Swift/UVOT and Hubble Space Telescope (HST) photometry \citep{Brown_etal_2010, Wang_etal_2012} and HST UV spectroscopy \citep{Foley_etal_2016}, though the effect is smaller than the NUV-blue/red differences \citep{Milne_etal_2013}.  The presence of the correlation in Swift photometry and HST photometry and spectrophotometry means it is not a side effect of the UVOT red leak.  To minimize the effect of such a correlation, we restrict our sample to SNe Ia with 1.0 $<$ \mb $<$ 1.4. We also exclude SNe Ia spectroscopically identified as peculiar subtypes (91T-like, Super-Chandrasekhar candidates, 91bg-like, or 2002cx-like).  All photometry comes from the Swift Optical/Ultraviolet Supernova Archive (SOUSA; \citealp{Brown_etal_2014_SOUSA}) and is available at the Swift SN website\footnote{http://swift.gsfc.nasa.gov/docs/swift/sne/swift\_sn.html} and the Open Supernova Archive\footnote{https://sne.space/} \citep{Guillochon_etal_2016}.  

\citet{Milne_etal_2013} found the NUV-blue/red dichotomy was apparent in the $uvw1-v$ and $u-v$ colors.  For our sample, many of the nearest and best-sampled SNe Ia (e.g. SNe~2005df, 2007af, 2011fe) were near or above the saturation limit in the UVOT optical ($u,b,v$) filters.  Ground-based Johnson $B$ and $V$-band observations can be substituted for Swift $b$ and $v$ with negligible differences \citep{Poole_etal_2008,Brown_etal_2009}.  The Swift $u,b,v$ filters are designated in lower-case to distinguish them from Johnson $U,B,V$ filters \citep{Poole_etal_2008}. The Swift/UVOT $u$ band, however, is broader and, uninhibited by atmospheric absorption or the poor blue response of CCDs, extends bluer than the Johnson $U$ or SDSS $u$ bands.  Ground-based $u$/$U$ bands effectively do not reach short enough wavelengths to see the spectroscopic differences identified in \citet{Milne_etal_2013} and \citet{Milne_etal_2015}.  \citet{Folatelli_etal_2010} found a $u-V$ color-width relation that is similar to what is seen in optical $B-V$ colors.  In this work we will focus on the $uvw1-v$ and $b-v$ colors, substituting ground-based $B$ and $V$-band measurements as needed, for the near-UV/optical color differences.  We separately examine the mid-UV using the uvm2 filter, as a further check on the reddening and intrinsic differences.  Mid-UV differences are found in SNe Ia with similar near-UV and optical properties \citep{Foley_Kirshner_2013}.

We fit the UVOT data around maximum light and up to 20 days after using the MLCS2k2 model \citep{Jha_etal_2007} in $b$ and $v$, the SN~2011fe $uvw1$ light curves \citet{Brown_etal_2012_11fe} for $uvw1$ and $u$, and the $uvm2$ and $uvw2$ curves of SN~2011fe for their respective filters.   From this stretched template we extract \mb~ and the magnitudes in the 6 UVOT bands interpolated at the time of the $B/b$-band maximum light using the respective best-fit light curves.  These parameters along with the host galaxy distances are listed in Table 1.  For the following SNe Ia we use $B$ and $V$ data from the listed references:  SNe 2005cf \citep{Pastorello_etal_2007,Wang_etal_2009_05cf}, 2005df \citep{Milne_etal_2010}, and 2011fe \citep{Richmond_Smith_2012}.

First we reproduce the $uvw1-v$ and $uvm2-uvw1$ temporal evolution of the NUV-blue and NUV-red classes from \citet{Milne_etal_2013}.  In the left panel of Figure \ref{fig_milne}, blue and red regions have been made by outlining the minimum and maximum colors at each epoch (binned by 5 days) within each group using the same SNe Ia.  Even within the groups, the $uvm2-uvw1$ colors have a large amount of scatter and there is a region occupied by both.
In  the right panel of Figure \ref{fig_milne} we show the new objects from our current study overplotted on those colored regions.

The Swift $uvw1$ filter does have an extended red tail reaching optical wavelengths (see \citealp{Brown_etal_2010}) which can complicate the direct translation of broadband photometry into monochromatic flux  \citep{Brown_etal_2016}.  For relative comparisons, the red tail does not affect the analysis.  If all SNe Ia have the same spectral shape, the colors and relative UV and optical contributions will be the same.  If there are spectral differences, the UV-brighter SNe Ia will have brighter UV magnitudes and bluer UV-optical colors and a smaller fractional contribution from the optical.  The change in the optical contribution is caused by the spectral differences rather than the red leak creating apparent differences not in the spectra.  Since we are studying apparent UV differences in optically-similar SNe Ia, the fact that some of the detected photons come from the optical cannot undermine the result.  The optical contribution can only dilute UV differences, as evidenced by the uvm2 filter showing more scatter in colors and absolute magnitudes than the uvw2 filter which has a bluer central wavelength but a long tail into the optical \citep{Brown_etal_2010}.  The forward-modeling technique we will use in comparing spectrophotometry from a reddened template with the photometry also allows us to naturally incorporate the red tail of the $uvw1$ filter for a robust comparison.

As a spectroscopic template we will use the bolometric spectral series of SN~2011fe from \citet{Pereira_etal_2013}, which incorporated HST UV spectra later published in \citet{Foley_Kirshner_2013} and \citet{Mazzali_etal_2014}.  SN~2011fe was a NUV-blue SN Ia \citep{Brown_etal_2012_11fe,Milne_etal_2013}.  So while it has become a commonly-used template object because of the extensive temporal and wavelength coverage of its observations, it is either part of a minority subclass of SNe Ia or near one extreme of a continuum.  Despite great efforts to broaden the parameter space of SNe Ia with UV spectroscopy (see e.g. \citealp{Foley_etal_2016}), we still lack a UV spectral series of a NUV-red SN Ia with a normal \mb.  So while we cannot test how well the photometry of NUV-red SNe Ia compares to a reddened NUV-red template, we can test whether their colors are consistent with that of a reddened NUV-blue.  Essentially, we are testing the hypothesis that SNe Ia with similar optical \mb~are intrinsically similar in the near-UV and that the observed differences are due to differences in reddening or reddening laws.  Within this context SN~2011fe is an excellent template to use.

\begin{figure} 
\resizebox{7.0cm}{!}{\includegraphics*{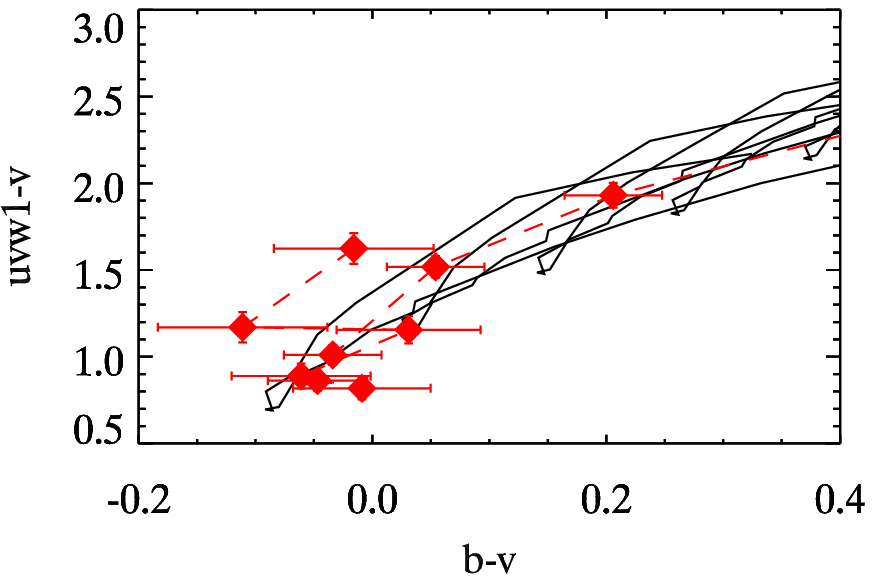}   }
\resizebox{7.0cm}{!}{\includegraphics*{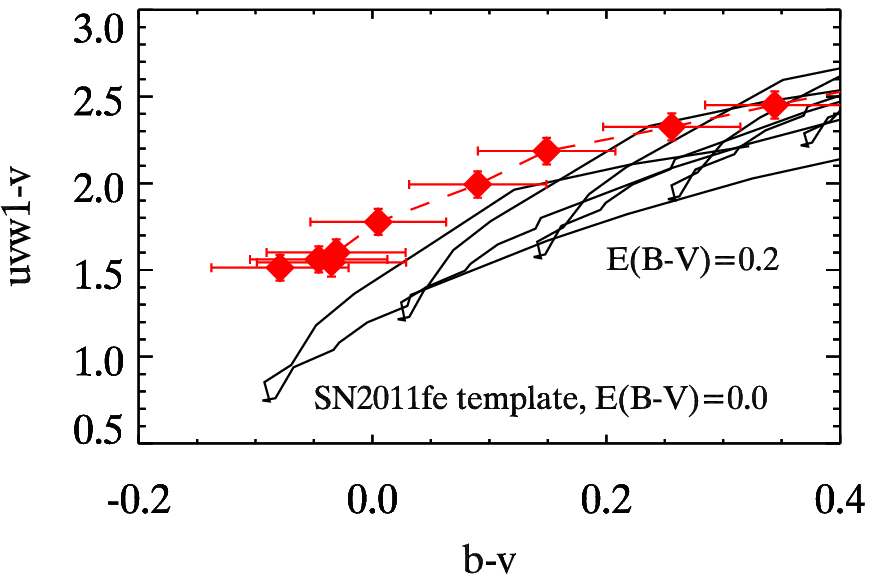}   }
\resizebox{7.0cm}{!}{\includegraphics*{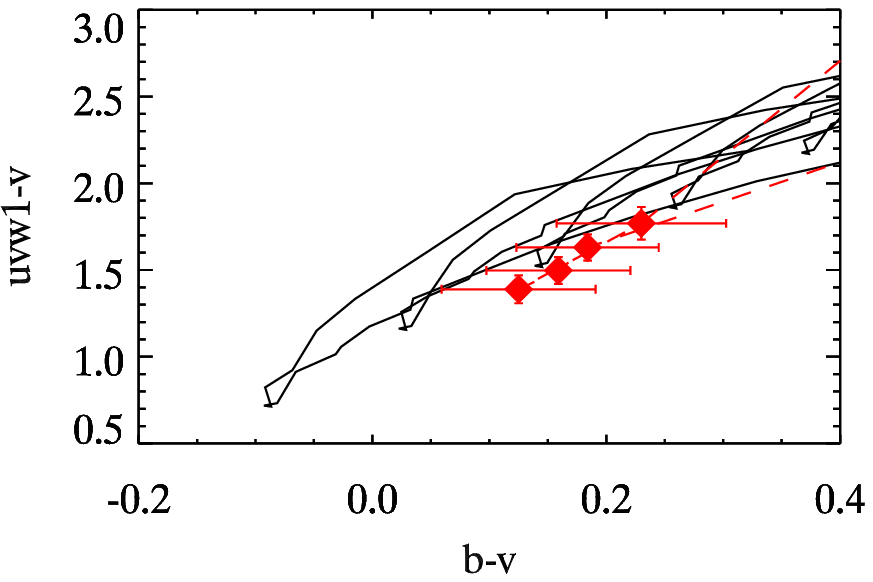}   }
\caption[Results]
        { Top Panel: $uvw1-v$ versus $b-v$ color evolution for SN2011by shown with red symbols.  The solid lines in each plot are the colors of SN~2011fe with different amounts of reddening applied (in steps of $E(B-V)=0.1$ mag). The colors of SN2011by are consistent with a SN~2011fe, confirming it is a NUV-blue SN Ia.
Middle Panel:  $uvw1-v$ versus $b-v$ color evolution for SN2012hr shown with red symbols.   The $uvw1-v$ colors are much redder than the templates reddened to comparable optical colors, indicating it is intrinsically a NUV-red SN Ia.

Bottom Panel:  $uvw1-v$ versus $b-v$ color evolution for SN2015F shown with red symbols.  The colors are consistent with a reddened version of SN~2011fe, suggesting it is intrinsically a NUV-blue SN Ia with approximately 0.2 mags of reddening.  The expected Milky Way reddening is $E(B-V)_{MW}=0.18$ mag \citep{Schlafly_Finkbeiner_2011}. } \label{fig_colors}    
\end{figure}

\section{Results} \label{results}

\subsection{Color Evolution} \label{colors}

 To predict the effect of reddening, we overplot the spectrophotometric color evolution of SN~2011fe with steps of 0.1 mag in $E(B-V)$ using a Milky Way (MW) reddening law (ie, R$_V$=3.1) from \citet{Cardelli_etal_1989}. This is shown in the left panel of Figure \ref{fig_redz}.  Using a different value for R$_V$ changes the offsets between the lines but not the temporal behavior.  
A reddening of $E(B-V)$=0.2 mag is sufficient to shift the $uvw1-v$ colors of SN~2011fe to the blue edge of the NUV-red group.  A reddening of  $E(B-V)=0.5$ mag would shift it through the entire range of observed colors for this sample. The post-maximum reddening in $uvw1-v$ has the same slope for the NUV-red and NUV-blue SNe Ia.  The timing of the pre-maximum minimum in the uvw1-v, however, is delayed for the NUV-red SNe Ia compared to the reddened color curves of SN~2011fe.

The observed $uvm2-uvw1$ colors have a large ($>$ 2 mag) scatter before maximum light, but the scatter decreases after maximum light as the colors generally become monotonically bluer.  The reddened curves required to match the spread in $uvw1-v$ colors would predict a much larger scatter in the post-maximum $uvm2-uvw1$ colors than is observed.  While some SNe Ia drop off of the plot because they become too faint in uvm2, the closer NUV-red objects would be expected to have a flat color evolution with very red colors, which we do not observe.  Despite the large scatter in uvm2 colors and behavior, it does not appear to be driven by reddening.

We also test for the effect of redshift on the observed colors.  Because of the steeply changing near-UV spectrum, the differences between the observed photometry and that which would be observed in the rest from of the source ($k$ corrections) are not negligible even for low redshifts (z$<$0.03) \citep{Brown_etal_2010}.   

The right panel of Figure \ref{fig_redz} shows the observer-frame spectrophotometric colors for the SN~2011fe template between z=0 and z=0.03 in steps of 0.01.  At a redshift of z=0.03, the $uvw1-v$ color is reddened by $\sim$0.3 mags.  This is not enough to make SN~2011fe look like a NUV-red SN, but it can contribute a significant shift.

\begin{figure} 
\resizebox{7.0cm}{!}{\includegraphics*{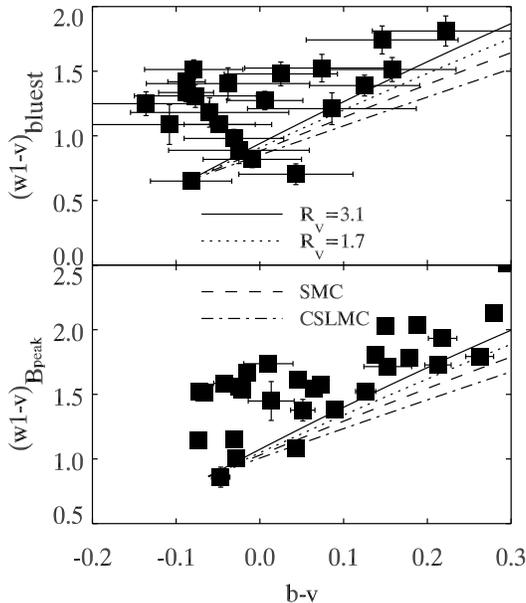}   }
\caption[Results]{  $uvw1-v$ versus $b-v$ colors of SNe Ia at the bluest epoch (top) and at the time of B-band maximum light (bottom).  The top panel has larger errors because they are directly from the single epoch photometry while the bottom points have been fit with a template.  Only SNe Ia with observations beginning five days before maximum light are included in these plots.  
 } \label{fig_colorcolor}    
\end{figure}

\subsection{Color-Color Evolution} \label{colorcolor}

While the bulk color evolution strongly suggests that reddening and redshift are not the dominant source of the color dispersion or the NUV-blue/red dichotomy, it does not tell us about individual objects.  To correct for the effects of redshift and reddening, one must have an accurate spectral model.  Lacking spectra for individual objects, we can use the spectral template of SN~2011fe and forward-model the effects of reddening and redshift on the observed magnitudes and colors.  

The NUV-optical and $b-v$ colors of SNe Ia become bluer until just before maximum light.  When plotted as $uvw1-v$ and $b-v$, the colors follow a narrow track to the blue corner and then retreat almost along the same track.  In Figure \ref{fig_colors} we show the color-color tracks of three SNe Ia along with that predicted from SN~2011fe (at the appropriate redshift) with different amounts of reddening.  We focus on the blue minimum of the colors as a pivot point, as epoch and reddening appear somewhat degenerate in the post-maximum phase.  In the top panel, the colors of SN~2011by are consistent with SN~2011fe.  We would thus classify SN~2011by as a NUV-blue SN.  In the middle panel, the colors of SN~2007af are vertically offset to the red from the track of the bluest colors.  We thus classify SN~2007af as NUV-red.
In the bottom panel, SN~2015F is consistent with SN~2011fe reddened by about $E(B-V)=0.2$ mag.  
According to the \citet{Schlafly_Finkbeiner_2011} reddening maps, the MW reddening along the line of sight amounts to $E(B-V)=0.18$ mag. 
Based solely on the color evolution, it would have been classified as a NUV-red SN.   We believe SN~2015F would more appropriately be labeled as a reddened NUV-blue.

We apply this procedure to all of the normal SNe Ia with observations beginning 5 days before the maximum light.  
Some SNe Ia are clearly offset vertically (in $uvw1-v$) compared to curves pivoting at comparable optical colors.  Several SNe Ia have colors comparable to the SN~2011fe model at zero or moderate reddening values.  Because we do not know the appropriate extinction law or how close SN~2011fe would be to an intrinsic color boundary, the line of separation between NUV-red and NUV-blue (especially reddened NUV-blue)  is still uncertain.  We can however, identify some individual SNe Ia which are consistent with SN~2011fe and a reddening law.  We can also identify some individual SNe Ia which have colors inconsistent with SN~2011fe and any of these reddening laws.

To compare multiple SNe Ia, Figure \ref{fig_colorcolor} plots the simultaneous $uvw1-v$ and $b-v$ colors at the bluest epoch and at the time of B-band maximum.  K-corrections have not been applied due to the limited number of spectral templates.  The effect of reddening is represented by taking the spectra from the SN~2011fe template series at its bluest $uvw1-v$ epoch and at the time of maximum light in the B band and reddening it with different extinction laws.  These include the \citet{Cardelli_etal_1989} parameterization with R$_V$=3.1 and 1.7, the SMC reddening law \citep{Prevot_etal_1984}, and a LMC reddening law modified by circumstellar scattering \citep{Goobar_2008}.  It is clear that SN~2015F and some of the other ``red'' SNe Ia are consistent with the reddened versions of SN~2011fe.

\subsection{UV-bright or optically-faint?} \label{absmags}

Having demonstrated near-UV color differences, it is natural to wonder whether NUV-blue SNe Ia are brighter in the near-UV or fainter in the optical.  In the analysis of \citet{Brown_etal_2010} we found that in the near-UV and optical, the scatter in the absolute magnitudes of the Swift SNe Ia sample was dominated by distance uncertainties.  That sample was small (8 SNe Ia in the \mb~range used here) and most are at small enough redshifts for peculiar velocities of the galaxies (after removing bulk flows from known overdensities as in e.g. \citealp{Mould_etal_2000}) to be a significant fraction of the recessional velocity.  The low redshift is a problem when the absence of redshift-independent distance estimates are unavailable and distances are computed from the recessional velocity.  Most of the Swift SNe Ia are still very nearby, but our sample is larger, includes more SNe Ia in the nearby Hubble flow (z$\gtrsim$ 0.02), and a growing number have Cepheid distances \citep{Riess_etal_2016}.  We are pursuing HST observations from which to obtain surface brightness fluctuation distances (SBF, \citealp{Tonry_etal_1997,Jensen_etal_2003,Jensen_etal_2015}) for many of our SNe Ia in late type galaxies.  The distances we use in this study are listed in Table 1.

Figure \ref{fig_abs} shows the $uvw1-v$ colors as well as the magnitudes in the $uvw1$ and $v$ bands (at the time of b-band maximum light) corrected for distance.  They are plotted with respect to the $b-v$ color (at $b$-band maximum light), which serves as a proxy for the amount of dust reddening.  For these plots we use the full sample of normals (i.e. we use the cut on \mb~but require only photometry at the time of $B$-band maximum rather than five days earlier).  We again show the expected track for a reddened version of SN~2011fe.  No extinction corrections have been applied, so the colors will be affected by a combination of Milky Way and host galaxy dust.

Of the five SNe Ia clearly blue in the top panel, SNe 2011fe and 2011ia are anomalously bright in the $uvw1$ filter, though not as bright as Super-Chandrasekhar candidates \citep{Brown_etal_2014, Scalzo_etal_2014}, SN~2004dt \citep{Wang_etal_2012}, or SN~2011de \citep{Brown_2014}. SN~2008hv is intermediate. SNe 2008Q and 2011by appear more normal in the UV.  SN~2008Q is anomalously faint in the optical, while SN~2011by is at the lower end of the optical brightness range, but not extremely so.  The difference between SNe 2011by and 2011fe is significant, as they have nearly identical spectral shapes in the near-UV and optical, but different mid-UV continuum levels and different optical absolute magnitudes  \citep{Foley_Kirshner_2013,Graham_etal_2015_twins}.  The absolute luminosity differences persist with the new Cepheid distances of \citep{Riess_etal_2016}.  An HST program (PI: Milne) will obtain SBF distances to the hosts of SNe 2008Q and 2008hv.

\section{Discussion} \label{discussion}

\begin{figure} 
\resizebox{7.0cm}{!}{\includegraphics*{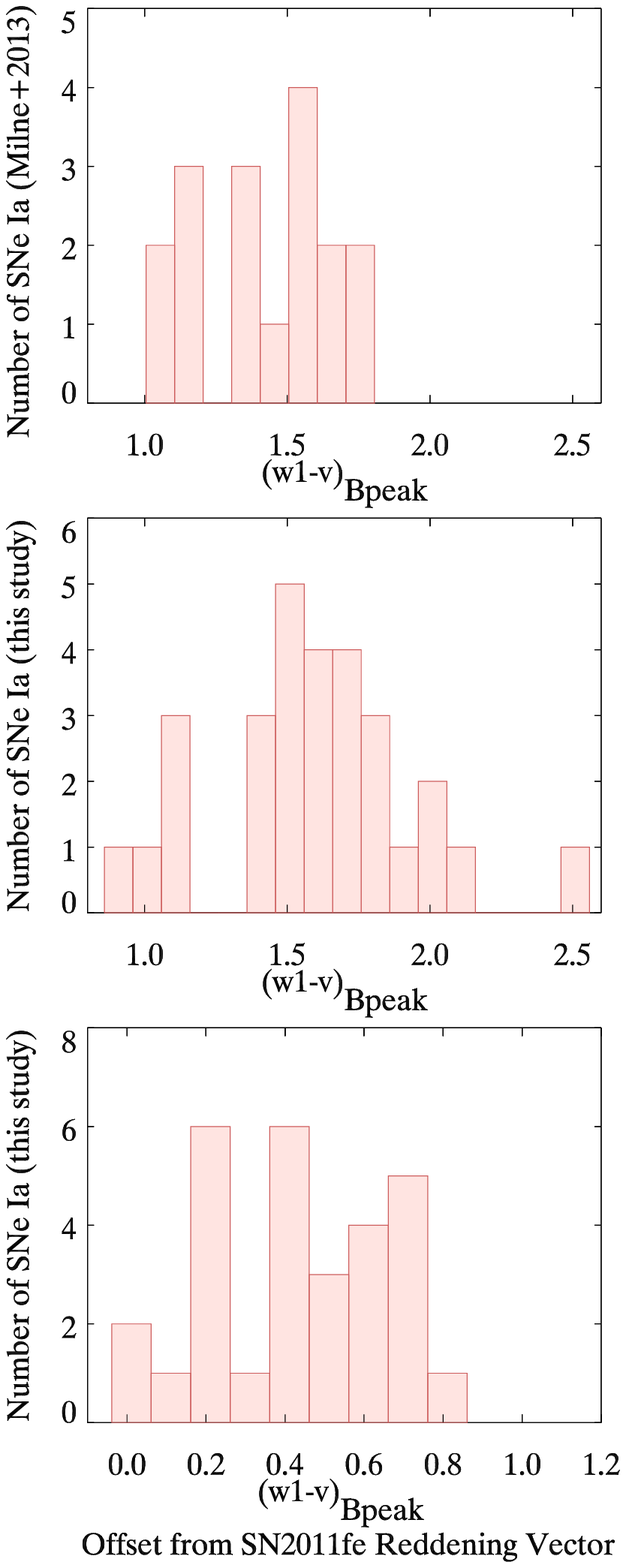}   }
\caption[Results]{ Top Panel: A histogram of the $uvw1-v$ colors (at the time of B-band maximum light) of the SNe Ia from \citet{Milne_etal_2013}. 
Middle Panel: A histogram of the $uvw1-v$ colors (at the time of B-band maximum light) of the SNe Ia from our sample with 1.0 $<$ \mb~ $<$ 1.4 and $(B-V)_{Bpeak} < $ 0.3.
Bottom Panel: A histogram of the vertical ($uvw1-v$) offset from the $uvw1-v$ versus $b-v$ colors of the reddened SN~2011fe model using the same sample as the middle panel.
} \label{fig_histograms}    
\end{figure}

Especially in the presence of reddening, it is difficult to determine whether the true distribution of intrinsic colors could be bimodal or just have a large dispersion.
In our sample, we still see a bimodality in the $uvw1-v$ colors similar to \citet{Milne_etal_2013}, as shown in Figure \ref{fig_histograms}.  However, our color-color evolution analysis also identifies SNe Ia (such as 2008ec and 2015F) which evolve like reddened NUV-blue SNe Ia.  When viewing the distributions with a diagonal cutoff roughly parallel to the (albeit divergent) reddening vectors, two distinct groups are no longer prominent.  We still don't see a continuous distribution, but the sample size, binning, and selection effects may play some role.  Variations in the reddening laws (which may or may not be related to the intrinsic colors of the SNe Ia) could also be mixing the distribution of modestly-reddened SNe Ia.  The ratio between the numbers of NUV-blue and red SNe Ia also depends on whether the colors are corrected for reddening and/or where the cutoff in $b-v$ is made for SNe Ia to be considered to have low reddening.

One concern raised in \citet{Milne_etal_2013} is that there could also be differences in the intrinsic optical $B-V$ colors from which reddening is determined, namely NUV-blue SNe Ia having bluer $B-V$ colors.  The expanded sample studied here includes SNe Ia which would be classified as NUV-red based on their $uvw1-v$ colors, but which have bluer $B-V$ colors than the NUV-blue SNe Ia.  We show here that the red cut off of the NUV-blue SNe Ia is the result of reddening.  A color excess of just $E(B-V)=0.2$ mag would cause SN~2011fe to be classified as NUV-red, limiting the $B-V$ color distribution of SNe Ia labeled NUV-blue. Fully understanding the intrinsic color distributions requires a better understanding of the multiwavelength reddening to the SNe Ia individually and an appropriate extrapolation of the reddening law to the UV.  An alternative route is a physical explanation for the observed differences, with the optical differences inferred from models which are constrained by the UV data.

\begin{figure*} 
\plottwo{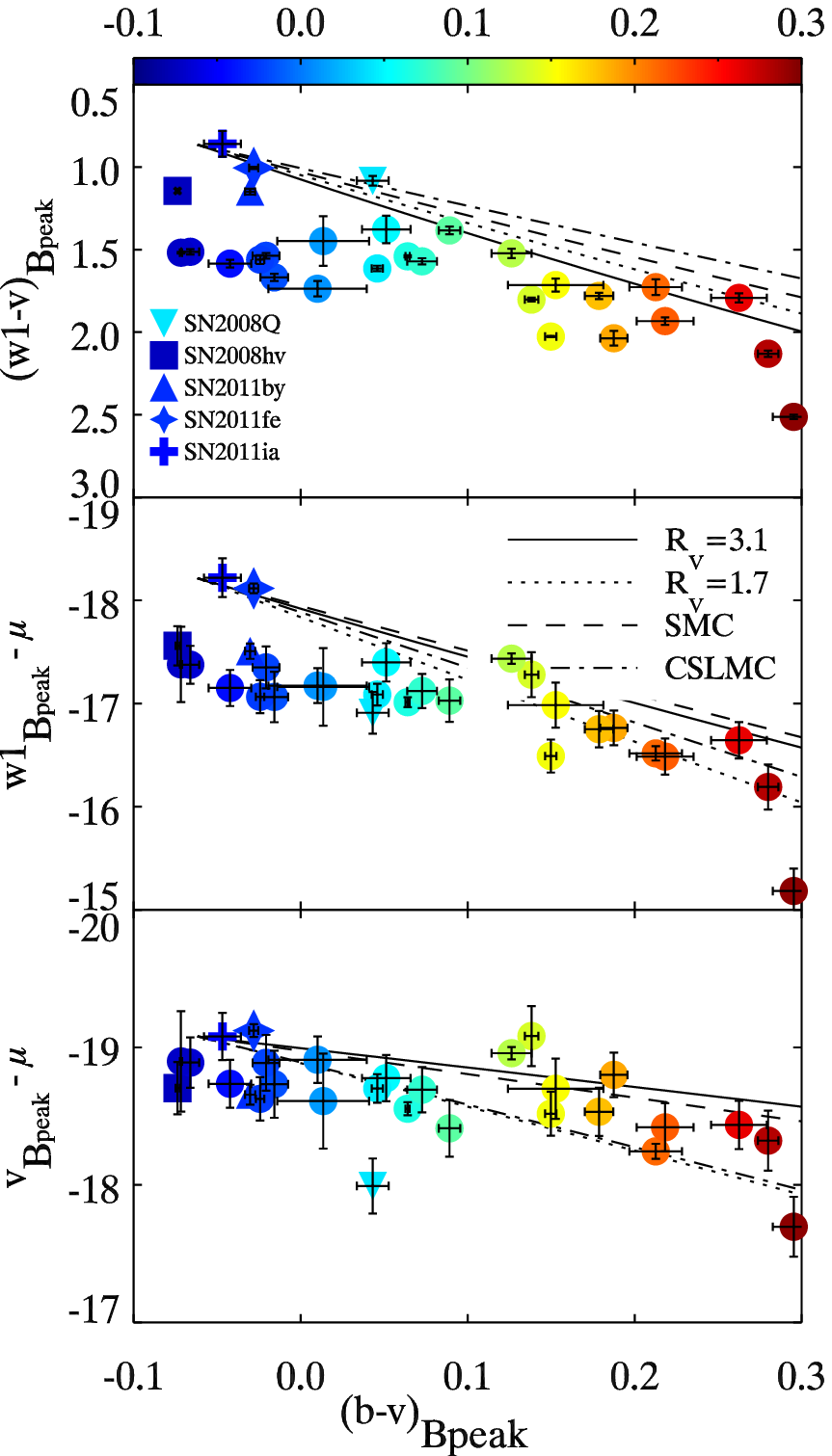}{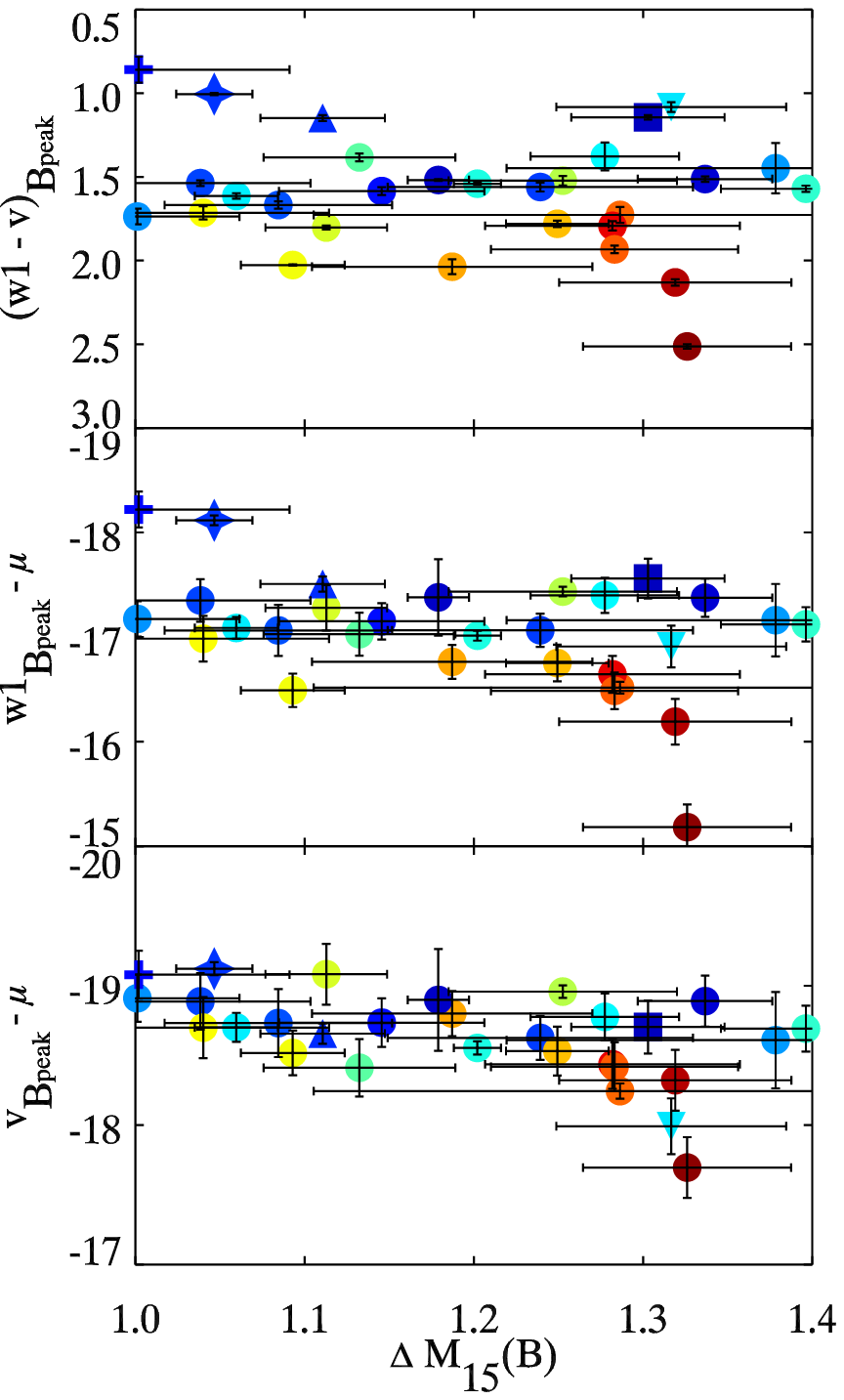} 
 \caption[Results]{ Left: $uvw1-v$ versus $b-v$ colors of SNe Ia  and the absolute magnitudes in $uvw1$ and v.  They are compared to the predicted tracks for SN~2011fe reddened with a MW extinction law.   
Right: $uvw1-v$ versus \mb~ of SNe Ia  and the absolute magnitudes in $uvw1$ and $v$.  The SNe Ia are color coded based on the $B-V$ color.  } \label{fig_abs}    
\end{figure*}

\citet{Milne_etal_2015} found similar near-UV color groupings in SNe Ia in Swift and HST SNe Ia at low redshifts (0.00 $\leq$ z$\leq$0.03), Keck spectrophotometry of SNe Ia at higher redshifts (0.18 $\leq$ z$\leq$0.53), and k-corrected $u-v$ colors from HST SNe Ia (z $\geq$ 0.6).  The population distribution, however, shifts with redshift; namely, the lower redshift SNe Ia are dominated by NUV-red SNe Ia while the higher (z$\sim$1 SNe Ia) are almost exclusively NUV-blue SNe Ia.  \citep{Cinabro_etal_2016} did not find evidence for a bimodality in the SDSS and SNLS samples (0.3 $\leq$ z$\leq$0.7), which had overlap with the spectroscopic samples used in \citet{Milne_etal_2015}.  More specifically, they found that their spectroscopic model designed to mimic the distributions seen in \citet{Milne_etal_2013} and \citet{Milne_etal_2015} predicted U-V colors which did not match the U-V from the SALT-II fits of the SDSS and SNLS SNe Ia.  Whether there are two distinct groups or a just a distribution in the colors, the observed shift in near-UV colors remains to be explained based on the SN properties or spectroscopic selection effects.
While selection effects are harder to model for the nearby SN Ia samples, it should be tested whether the models for spectral variation \citep{Betoule_etal_2014,Kessler_etal_2013,Scolnic_Kessler_2016} can match the observed local distribution shown here and extreme cases such as SN~2011fe.
As shown in Figures \ref{fig_abs} and \ref{fig_colorcolor}, there is UV variation that is not correlated with the optical colors.  Thus a color correction which combines intrinsic color variations and reddening together will not be accurate in the UV.


\section{Summary}\label{summary} 

We have used an expanded sample of  29 SNe Ia with 1.0 $<$ \mb~$<$ 1.4 to study the effects of reddening and redshift on the observed near-UV colors of SNe Ia.

$\bullet$ SNe Ia with similar optical colors have a 0.7 mag spread in $uvw1-v$ colors, which cannot be caused by known extinction laws.

$\bullet$ The $uvw1-v$ color evolution can be broadly reproduced by reddening the SN~2011fe template, but the $uvm2-uvw1$ colors and color evolution are not consistent with the same amount of reddening.  

$\bullet$ The redshift range of the Swift SNe Ia (0.00 $\leq$ z$\leq$0.03) results in a  $\sim$0.3 mag difference in $uvw1-v$ using the SN~2011fe template.  However, as shown in \citet{Milne_etal_2013}, redshift does not drive the color differences  (over 80\% of our SNe Ia have z$<$0.02 where the difference is 0.2 mag or less) and different UV spectra give different k-corrections.

$\bullet$ This larger sample also shows a dichotomy of $uvw1-v$ colors.  Many, but not all, of the red SNe Ia are consistent with the colors of a reddened SN~2011fe model.   At least six SNe Ia have color-color evolution shifted to the red in the $uvw1-v$ colors but not in the $b-v$ colors, clearly not the effect of reddening, while optically-red SNe Ia are more difficult to interpret.  

$\bullet$ Out of five SNe Ia with blue $uvw1-v$ colors and low reddening, two are brighter than normal in the UV and three are at the faint end of normal in the optical.

$\bullet$ SN~2011fe, often considered to be the best-observed normal SN Ia, does not appear to be normal in the UV, but is overluminous by $\sim$0.8 mags in $uvw1$ compared to minimally-reddened SNe Ia with comparable \mb.

\vspace{10mm}

\section{Acknowledgements}\label{acknowledgements} 

P. J. B. and the Swift Optical/Ultraviolet Supernova Archive (SOUSA) are supported by NASA's Astrophysics Data Analysis Program through grant NNX13AF35G.
We acknowledge the use of Swift data obtained under the agreement with 
the Instrument Center for Danish Astrophysics (IDA).
M. D. S. is funded by generous support provided by the Danish Agency for Science and Technology and Innovation realized through a Sapere Aude Level 2 grant and from the Villum Foundation. 

%
%
%
%
%
%
%
%
%
%
%
%
%
\clearpage



\begin{deluxetable}{llllrll}




\tablecaption{Supernova Information}

\tablenum{1}

\tablehead{\colhead{SN name} & \colhead{\mb} & \colhead{w1$_{Bpeak}$} & \colhead{V$_{Bpeak}$}  & \colhead{(B$-$V)$_{Bpeak}$}  & \colhead{Distance} & \colhead{Reference} \\ 
\colhead{} & \colhead{(mag)} & \colhead{(mag)} & \colhead{(mag)} & \colhead{(mag)}  & \colhead{(mag)}  & \colhead{ } } 

\startdata
SN~2015F  & 1.29 $\pm$ 0.18 & 14.994 $\pm$  0.044 & 13.267 $\pm$  0.016 &  0.213 $\pm$  0.016 & 31.511   $\pm$  0.053 & Cepheids \citep{Riess_etal_2016} \\
SN~2013gy & 1.34 $\pm$ 0.04 & 16.456 $\pm$  0.015 & 14.943 $\pm$  0.005 & -0.066 $\pm$  0.005 & 33.832   $\pm$  0.183 & $z$ \citep{2005AJ....130.1037C} \\
SN~2013gs & 1.15 $\pm$ 0.06 & 17.085 $\pm$  0.019 & 15.500 $\pm$  0.013 & -0.042 $\pm$  0.013 & 34.235   $\pm$  0.174 & $z$ \citep{1999PASP..111..438F} \\
SN~2013ex & 1.13 $\pm$ 0.06 & 16.163 $\pm$  0.023 & 14.780 $\pm$  0.006 &  0.089 $\pm$  0.007 & 33.191  $\pm$   0.205 & $z$ \citep{1991rc3..book.....D} \\
SN~2013eu & 1.33 $\pm$ 0.06 & 17.749 $\pm$  0.005 & 15.236 $\pm$  0.013 &  0.295 $\pm$  0.013 & 32.932  $\pm$   0.218 & $z$ \citep{1999PASP..111..438F} \\
SN~2013cs & 1.11 $\pm$ 0.04 & 15.668 $\pm$  0.012 & 13.867 $\pm$  0.006 &  0.128 $\pm$  0.006 & 32.927  $\pm$   0.217 & $z$ \citep{2011ApJS..197...28P} \\
SN~2012ht & 1.25 $\pm$ 0.07 & 14.475 $\pm$  0.026 & 12.952 $\pm$  0.012 &  0.126 $\pm$  0.012 & 31.910  $\pm$   0.043 & Cepheids \citep{Riess_etal_2016} \\
SN~2012hr & 1.08 $\pm$ 0.07 & 15.427 $\pm$  0.020 & 13.759 $\pm$  0.008 & -0.016 $\pm$  0.008 & 32.491  $\pm$   0.244 & $z$ \citep{2004AJ....128...16K} \\
SN~2011im & 1.28 $\pm$ 0.08 & 17.505 $\pm$  0.022 & 15.712 $\pm$  0.017 &  0.262 $\pm$  0.017 & 34.149  $\pm$   0.175 & $z$ \citep{2005AJ....130.1037C} \\
SN~2011ia & 1.00 $\pm$ 0.09 & 16.129 $\pm$  0.077 & 15.269 $\pm$  0.011 & -0.047 $\pm$  0.011 & 34.348  $\pm$   0.172 & $z$ \citep{2012ApJS..199...26H} \\
SN~2011fe & 1.05 $\pm$ 0.02 & 11.023 $\pm$  0.006 & 10.017 $\pm$  0.003 & -0.028 $\pm$  0.003 & 29.139  $\pm$   0.046 & Cepheids \citep{Riess_etal_2016} \\
SN~2011by & 1.11 $\pm$ 0.04 & 14.081 $\pm$  0.018 & 12.932 $\pm$  0.003 & -0.030 $\pm$  0.003 & 31.588  $\pm$   0.072 & Cepheids \citep{Riess_etal_2016} \\
SN~2011ao & 1.04 $\pm$ 0.07 & 15.896 $\pm$  0.014 & 14.358 $\pm$  0.008 & -0.021 $\pm$  0.008 & 33.246  $\pm$   0.203 & $z$ \citep{2007ApJS..172..634A} \\
SN~2011B  & 1.38 $\pm$ 0.16 & 14.283 $\pm$  0.148 & 12.834 $\pm$  0.027 &  0.014 $\pm$  0.027 & 31.444  $\pm$   0.345 & $z$ \citep{2006AJ....131..185R} \\
SN~2010kg & 1.28 $\pm$ 0.07 & 17.718 $\pm$  0.016 & 15.785 $\pm$  0.017 &  0.218 $\pm$  0.017 & 34.203  $\pm$   0.174 & $z$ \citep{1991rc3..book.....D} \\
SN~2010gp & 1.19 $\pm$ 0.08 & 18.278 $\pm$  0.044 & 16.241 $\pm$  0.008 &  0.188 $\pm$  0.008 & 35.042  $\pm$   0.162 & $z$\citep{1993ApJ...414L..13D} \\
SN~2010gn & 1.24 $\pm$ 0.09 & 18.489 $\pm$  0.026 & 16.929 $\pm$  0.002 & -0.024 $\pm$  0.003 & 35.554  $\pm$   0.158 & $z$ \citep{2014MNRAS.438.1391P} \\
SN~2010ev & 1.32 $\pm$ 0.07 & 16.727 $\pm$  0.018 & 14.597 $\pm$  0.006 &  0.280 $\pm$  0.006 & 32.919  $\pm$   0.218 & $z$ \citep{2005MNRAS.361...34D} \\
SN~2009cz & 1.00 $\pm$ 0.06 & 17.549 $\pm$  0.037 & 15.813 $\pm$  0.029 &  0.010 $\pm$  0.029 & 34.723  $\pm$   0.166 & $z$ \citep{2004ApJ...607..202M} \\
SN~2009Y  & 1.04 $\pm$ 0.07 & 15.968 $\pm$  0.026 & 14.253 $\pm$  0.029 &  0.153 $\pm$  0.029 & 32.952  $\pm$   0.217 & $z$ \citep{2005AJ....130.1037C} \\
SN~2008hv & 1.30 $\pm$ 0.05 & 16.031 $\pm$  0.013 & 14.887 $\pm$  0.001 & -0.074 $\pm$  0.001 & 33.590  $\pm$   0.190 & $z$ \citep{2003AJ....126.2268W} \\
SN~2008ec & 1.25 $\pm$ 0.03 & 17.409 $\pm$  0.017 & 15.629 $\pm$  0.009 &  0.179 $\pm$  0.009 & 34.161  $\pm$   0.175 & $z$ \citep{1996ApJS..106...27K} \\
SN~2008Q  & 1.32 $\pm$ 0.07 & 14.830 $\pm$  0.028 & 13.748 $\pm$  0.009 &  0.043 $\pm$  0.009 & 31.740  $\pm$   0.200 & SBF \citep{Jensen_etal_2003} \\
SN~2007co & 1.09 $\pm$ 0.03 & 18.760 $\pm$  0.001 & 16.734 $\pm$  0.003 &  0.150 $\pm$  0.003 & 35.251  $\pm$   0.160 & $z$ \citep{1996AJ....112.1803M} \\
SN~2007af & 1.20 $\pm$ 0.01 & 14.760 $\pm$  0.007 & 13.219 $\pm$  0.001 &  0.064 $\pm$  0.001 & 31.773  $\pm$   0.047 & Cepheids \citep{Riess_etal_2016} \\
SN~2006ej & 1.28 $\pm$ 0.04 & 17.252 $\pm$  0.082 & 15.874 $\pm$  0.015 &  0.051 $\pm$  0.015 & 34.651  $\pm$   0.167 & $z$ \citep{Abazajian_etal_2003} \\
SN~2006dm & 1.40 $\pm$ 0.05 & 17.690 $\pm$  0.016 & 16.119 $\pm$  0.009 &  0.073 $\pm$  0.009 & 34.811  $\pm$   0.165 & $z$ \citep{1998AandAS..130..333T} \\
SN~2005df & 1.18 $\pm$ 0.02 & 13.910 $\pm$  0.005 & 12.392 $\pm$  0.001 & -0.072 $\pm$  0.001 & 31.290  $\pm$   0.366 & $z$ \citep{2004AJ....128...16K} \\
SN~2005cf & 1.06 $\pm$ 0.02 & 15.174 $\pm$  0.016 & 13.560 $\pm$  0.003 &  0.046 $\pm$  0.003 & 32.261  $\pm$   0.104 & Cepheids \citep{Riess_etal_2016} \\

\enddata


\tablecomments{z -- redshift distance is derived using the referenced recessional velocity corrected for the galaxy-flow model of \citet{Mould_etal_2000} implemented by the NASA Extragalactic Database (NED) and converted to a distance modulus assuming H$_0$=72 km s$^{-1}$ Mpc$^{-1}$ \citep{Freedman_etal_2001}}


\end{deluxetable}

\bibliographystyle{apj}

\bibliography{bibtex}

\end{document}